\documentclass[a4paper,11pt]{article}
\pdfoutput=1 

\usepackage{jheppub_modified} 

\usepackage[T1]{fontenc} 

\usepackage{verbatim}
\usepackage{color}
\usepackage{graphicx}
\usepackage{subcaption}

\usepackage{bigcenter}

\usepackage{sectsty}
\allsectionsfont{\boldmath}

\usepackage{xspace}
\usepackage[dvipsnames,table]{xcolor}
\usepackage{tabularx}
\usepackage[normalem]{ulem}
\newcolumntype{C}[1]{>{\centering\arraybackslash}p{#1}}

\newcommand{\eq}[1]{Eq.~(\ref{#1})}
\newcommand{\bib}[1]{Ref.~\cite{#1}}

\newcommand{\fig}[1]{Fig.~\ref{#1}}
\newcommand{\tab}[1]{Table~\ref{#1}}

\newcommand{\bea}{\begin{eqnarray}}
\newcommand{\eea}{\end{eqnarray}}

\newcommand{\eps}{\epsilon}

\newcommand{\crn}{\nonumber \\}

\newcommand{\fr}{\frac}

\newcommand{\gev}{{\unskip\,\text{GeV}}}
\newcommand{\tev}{{\unskip\,\text{TeV}}}

\title{Joint polarizations of W pair production at the LHC at NLO QCD+EW accuracy}

\author[a]{Thi Nhung Dao,}
\author[a]{\underline{Duc Ninh Le}}

\affiliation[a]{Phenikaa Institute for Advanced Study, Phenikaa University, Hanoi 12116, Vietnam}

\emailAdd{nhung.daothi@phenikaa-uni.edu.vn}
\emailAdd{ninh.leduc@phenikaa-uni.edu.vn}

\abstract{In this contribution, we present new results of next-to-leading order (NLO) electroweak corrections to the doubly polarized $W^+W^-$ production cross sections at the LHC, via the full leptonic final state. This calculation has been recently achieved independently by two groups: one in Germany and our group in Vietnam. A comparison of the two results will be presented. We include also the NLO QCD corrections in the numerical analysis since they are dominant and therefore important for comparison with experimental results. New results of integrated cross section 
for future proton-proton colliders with $\sqrt{s}=$ $27\tev$, $50\tev$, $100\tev$ are provided. 
Moreover, a detailed explanation of the $\sigma_\text{TT}>\sigma_\text{LT}>\sigma_\text{LL}$ hierarchy based on the Born approximation is given.
}

\note{This is an extended version of the talk presented at the 49th Vietnam Conference on Theoretical Physics, Hue, Vietnam (2024).}

\begin{document}
\maketitle
\flushbottom

\section{Introduction}
\label{sect:intro}
Polarization of a massive gauge boson has attracted attention since the discovery of the Standard Model (SM) and 
the subsequent observation of the $W^\pm$ and $Z$ bosons at the CERN proton-antiproton collider SPS (Super Proton Synchrotron) in 1983. 

Since then, a lot of theoretical and experimental works have been performed. Notably, the polarization fractions for $e^+e^- \to W^+W^-$ process have been measured at LEP by the OPAL \cite{OPAL:2000wbs} and DELPHI \cite{DELPHI:2009wdg} collaborations. 
More recently, at the LHC, joint-polarized cross sections in $W^\pm Z$, $ZZ$, and 
same-sign $W^\pm W^\pm jj$ productions have been reported in 
\cite{ATLAS:2022oge,ATLAS:2024qbd} (ATLAS), \cite{ATLAS:2023zrv} (ATLAS), and \cite{CMS:2020etf} (CMS), respectively. 

On the theoretical side, the recent 
next-to-leading order (NLO) QCD and electroweak (EW) predictions for doubly-polarized cross sections 
have been provided for $ZZ$ \cite{Denner:2021csi}, $W^\pm Z$ \cite{Denner:2020eck,Le:2022lrp,Le:2022ppa}, 
$W^+ W^-$ \cite{Denner:2020bcz,Denner:2023ehn,Dao:2023kwc,Dao:2024ffg}, for fully leptonic decays. 
Next-to-next-to-leading order (NNLO) QCD results have been obtained for $W^+W^-$ \cite{Poncelet:2021jmj}. 
Semileptonic final state has been considered in \cite{Denner:2022riz} for the case of $WZ$ at NLO QCD, in \cite{Denner:2024ndl} for 
triboson production at NLO QCD+EW, and in \cite{Denner:2024xul} for vector boson scattering at leading order (LO). 

Going beyond the fixed order, new results in \cite{Hoppe:2023uux} show that 
it is now possible to simulate polarized events, for multi-boson production processes, 
at the precision level of approximate fixed-order NLO QCD corrections
matched with parton shower using the Monte-Carlo generator SHERPA. 
In addition, the above full NLO QCD calculations in 
$ZZ$ \cite{Denner:2021csi}, $W^\pm Z$ \cite{Denner:2020eck}, 
$W^+ W^-$ \cite{Denner:2020bcz} have been implemented in the 
POWHEG-BOX framework \cite{Pelliccioli:2023zpd}, thereby incorporating parton-shower effects. 
Very recently, polarized $ZZ$ pairs via gluon fusion 
have been generated using the combination of FeynRules and MadGraph5{\_}aMC@NLO \cite{Javurkova:2024bwa}, 
allowing for another option of realistic simulation. 

In this contribution, we report on the new finding of \bib{Dao:2023kwc} for the $pp \to W^+W^- \to e^+\nu_e \mu^-\bar{\nu}_\mu$ process 
at NLO QCD+EW and perform a comparison to 
the NLO EW results of \bib{Denner:2023ehn} which was published on arXiv one day before ours. 
The content of this report is significantly improved in comparison to the talk presented at the conference in the following respects: (i) the references are updated to include a few new relevant articles which have been published since the talk, (ii) 
a few comments are added to make connection to our new publication \cite{Dao:2024ffg}, 
(iii) a detailed explanation of the $\sigma_\text{TT}>\sigma_\text{LT}>\sigma_\text{LL}$ hierarchy based on the Born approximation is provided, 
(iv) integrated cross sections for future proton-proton colliders with $\sqrt{s}=$ $27\tev$, $50\tev$, $100\tev$ 
are given.   
Other numerical results for the case of $13\tev$ are the same as those presented at the talk. 
\section{Calculation method}
\label{sect:cal}
In order to separate the different polarization contributions of the $W^+W^-$ system, we use the 
Double Pole Approximation (DPA) \cite{Aeppli:1993cb,Aeppli:1993rs,Denner:2000bj} which works in three steps: 
\begin{itemize}
\item Select all diagrams with 2 $s$-channel resonances: $W^+\to e^+ \nu_e$, $W^-\to \mu^- \bar{\nu}_\mu$;
\item Factorize the amplitude into production and decay parts, and replace the propagators of 
the resonances by the Breit-Wigner formulae;
\item Apply an on-shell projection on the momenta of the production process and on the momenta of the decays to make 
each production and decay amplitudes gauge invariant.  
\end{itemize}
\begin{figure}[ht!]
  \centering
  \includegraphics[width=0.8\textwidth]{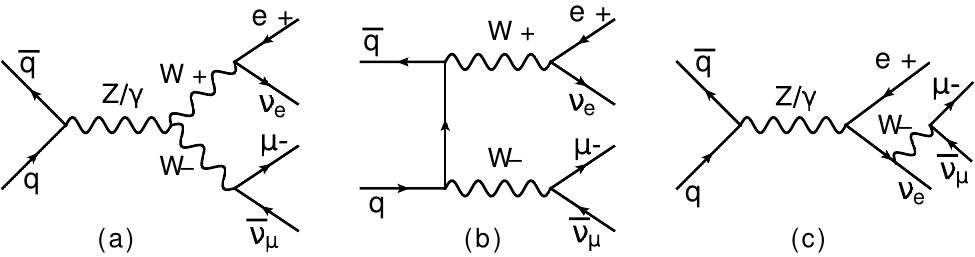}
  \caption{Representative Feynman diagrams at leading order. Diagrams (a) and 
  (b) are doubly resonant hence included in the DPA, while (c) is excluded.}
  \label{fig:LO_diags}
\end{figure}
The first step can be better understood by looking at the leading order diagrams in \fig{fig:LO_diags}. 
The doubly resonant diagrams (a) and (b) are selected for the DPA while the singly resonant diagram (c) is excluded.   

In the second step, the unpolarized amplitude is then factorized at leading order as follows (writing $V_1 = W^+$, $V_2 = W^-$, 
$l_1 = e^+$, $l_2 = \nu_e$, $l_3 = \mu^-$, $l_4 = \bar{\nu}_\mu$):
\bea
\mathcal{A}_\text{LO,DPA}^{\bar{q}q\to V_1V_2\to 4l} = \fr{1}{Q_1Q_2}
\sum_{\lambda_1,\lambda_2=1}^{3}
\mathcal{A}_\text{LO}^{\bar{q}q\to V_1V_2}(\hat{k}_i,\lambda_1,\lambda_2)\mathcal{A}_\text{LO}^{V_1\to
    l_1l_2}(\hat{k}_i,\lambda_1)\mathcal{A}_\text{LO}^{V_2\to l_3l_4}(\hat{k}_i,\lambda_2)
,\label{eq:LO_DPA}
\eea
with 
\bea
Q_j = q_j^2 - M_{V_j}^2 + iM_{V_j}\Gamma_{V_j}\;\; (j=1,2),
\label{eq:Qi_def}
\eea
where $q_1 = k_3+k_4$, $q_2 = k_5 + k_6$, $M_{V_j}$ and $\Gamma_{V_j}$ are the
physical mass and width of the gauge boson $V_j$, and $\lambda_j$ are the
polarization indices of the gauge bosons. Notice that, different from the narrow width approximation, 
the off-shell momenta are kept in the denominator of the gauge boson propagators to better describe the 
two resonances. 

For \eq{eq:LO_DPA} to be gauge invariant, we have to make each amplitude factor in the r.h.s. gauge invariant. 
This is achieved by requiring that the momentum set of each amplitude is on-shell, meaning they satisfy the 
condition $\hat{k}_i^2 = m_i^2$ with $m_i$ being the physical mass of the $i$-th particle of the production or decay process. 
To distinguish the on-shell momenta, we denote them with a hat. 

In practice, the on-shell momenta $\hat{k}_i$ are calculated from the off-shell momenta $k_i$ using an on-shell mapping. 
This mapping is not unique and different mappings give different results. 
The differences are very small, of order $\alpha \Gamma_V/(\pi M_V)$ \cite{Denner:2000bj}, being the intrinsic 
uncertainty of the DPA due to a finite width. 
In the present work, we use the LO on-shell mapping $\text{DPA}^{(2,2)}$ as defined in \cite{Denner:2021csi}. 
This completes the description of the third step mentioned above. 

The above procedure can be extended for NLO QCD and EW corrections. 
The Catani-Seymour subtraction method \cite{Catani:1996vz} is very useful and applied here. The massive dipoles are provided in \cite{Catani:2002hc} for dimensional regularization 
and in \cite{Dittmaier:1999mb} for mass regularization for production processes. 
The dipole subtraction method for decay processes 
was developed in \cite{Basso:2015gca}. 
We need the massive dipoles because both the production and 
decay processes have massive gauge bosons in the external legs. 
Note that, even though the massive dipole terms 
provided in \cite{Catani:2002hc,Dittmaier:1999mb,Basso:2015gca} are for fermions, they 
can be used for gauge bosons as well, because the soft singularity structure is identical 
for fermions and bosons (see e.g. Eq. (7.13) of \cite{Denner:1991kt}) and the collinear singularity 
(which is spin dependent) does not cause problems for the $W$ bosons due to their large mass compared to the collision energy at the LHC. 

The implementation of the Catani-Seymour method in the DPA is nontrivial because we have to deal with two kinds of mappings, the on-shell mappings (on-shell limits) and the Catani-Seymour mappings (soft and collinear limits), which do not commute. 
The order to apply them was specified in \cite{Le:2022ppa,Dao:2023kwc}, see also \cite{Denner:2021csi,Denner:2023ehn}. 
In this work, we follow the method of \cite{Le:2022ppa,Dao:2023kwc}, where the reader can find all the calculation details. 

It is important to note that the separation of the unpolarized amplitude into a sum of polarized amplitudes as 
in \eq{eq:LO_DPA} must be done for all NLO EW correction processes: virtual, photon radiation, quark-photon induced processes  
(and similarly for NLO QCD corrections). In this way, we can define polarized cross sections at NLO by summing up all the corrections 
for a given polarization (e.g. selecting $\lambda_1=\lambda_2=2$ for the LL polarization). 

Since each massive gauge boson has three polarization states, two transverse modes ($\lambda=1,3$ in \eq{eq:LO_DPA}) 
and one longitudinal mode ($\lambda=2$ in \eq{eq:LO_DPA}); the unpolarized cross section, being proportional 
to the unpolarized amplitude squared, can be separated into five terms:
\bea
\sigma_\text{unpol} = \sigma_\text{LL} + \sigma_\text{LT} + \sigma_\text{TL} + \sigma_\text{TT} + \sigma_\text{interf},
\eea
where $\sigma_\text{LL} \propto |\mathcal{A}_{22}|^2$, 
$\sigma_\text{LT} \propto |\mathcal{A}_{21}+\mathcal{A}_{23}|^2$ (a coherent sum of the $\mathcal{A}_{21}$ and $\mathcal{A}_{23}$ amplitudes), 
$\sigma_\text{TL} \propto |\mathcal{A}_{12}+\mathcal{A}_{32}|^2$, 
$\sigma_\text{TT} \propto |\mathcal{A}_{11}+\mathcal{A}_{13}+\mathcal{A}_{31}+\mathcal{A}_{33}|^2$. 
The last term is the interference between the LL, LT, TL, and TT amplitudes. 
This interference term vanishes if the integration over the full phase space of the decay products is performed \cite{Denner:2020bcz}. 
This is however not the case in realistic calculations as kinematic cuts are applied to the decay products, leading to a non-vanishing 
interference.
 
We note that while the unpolarized cross section is Lorentz invariant the individual polarized cross sections in the 
r.h.s. are not Lorentz invariant. Their values therefore are reference frame dependent. 
Various choices can be made, including the Laboratory frame, the partonic frame, or the $VV$ center-of-mass frame. 
In this work, we will choose the $VV$ frame, as it is the most natural choice to study the polarization of a diboson system.  

We now come to our computer tools. 
The numerical results of this work are obtained using our in-house computer program {\tt MulBos} (MultiBoson production), which 
has been used for our previous papers \cite{Le:2022lrp,Le:2022ppa,Dao:2023kwc,Dao:2024ffg}. 
The ingredients of this program include the helicity amplitudes for the production and decay processes, generated by 
{\tt FeynArt} \cite{Hahn:2000kx} and {\tt FormCalc} \cite{Hahn:1998yk}, an in-house library for one-loop integrals named {\tt LoopInts}. 
The tensor one-loop integrals are calculated using the standard technique of Passarino-Veltman reduction \cite{Passarino:1978jh}, 
while the scalar integrals are computed as in \cite{'tHooft:1978xw, Nhung:2009pm, Denner:2010tr}.  
The phase space integration is done using the Monte-Carlo integrator {\tt BASES} \cite{Kawabata:1995th}, with the help of 
useful resonance mapping routines publicly available in {\tt VBFNLO} \cite{Baglio:2024gyp}. 
Our code has been carefully checked by making sure that all UV and IR divergences cancel and singular limits of the dipole 
subtraction terms behave correctly. Comparisons with the results of \cite{Denner:2020bcz,Denner:2020eck,Denner:2021csi,Denner:2023ehn} 
have been done for all $ZZ$, $WZ$ and $W^+W^-$ processes, showing good agreements. 
The largest discrepancy is at the level of $0.4\%$ for the $W^+W^-$ case, which will be discussed more later.
\section{Numerical results}
\label{sect:res}
Using the ATLAS setup of (with $\ell=e,\mu$, missing energy is due to the neutrinos)
\begin{align}
        & p_{T,\ell} > 27\gev, \quad p_{T,\text{miss}} > 20\gev, \quad |\eta_\ell|<2.5, \quad m_{e\mu} > 55\gev,\crn
        & \text{jet veto (no jets with $p_{T,j}>35\gev$ and $|\eta_j|<4.5$)},
\end{align}
where the jet veto is used to suppress the top-quark and other QCD backgrounds, we obtain the following 
results for the integrated polarized cross sections, see \tab{tab:xs_fr}.
\begin{table}[h!]
 \renewcommand{\arraystretch}{1.3}
\begin{bigcenter}
\setlength\tabcolsep{0.03cm}
\fontsize{8.0}{8.0}
\begin{tabular}{|c|c|c|c|c|c|c|c|c|c|}\hline
  & $\sigma_\text{LO}\,\text{[fb]}$ & $\sigma^\text{QCD}_\text{NLO}\,\text{[fb]}$ & $\sigma^\text{QCDEW}_\text{NLO}\,\text{[fb]}$ & $\sigma_\text{all}\,\text{[fb]}$ & $\overline{\delta}_\text{EW}\,\text{[\%]}$ & $\overline{\delta}_{gg}\,\text{[\%]}$ & $\overline{\delta}_{b\overline{b}}\,\text{[\%]}$ & $\overline{\delta}_{\gamma\gamma}\,\text{[\%]}$ & $f_\text{all}\,\text{[\%]}$ \\
\hline
{\fontsize{7.0}{7.0}$\text{Unpolarized}$} & $198.14(1)_{-6.5\%}^{+5.3\%}$ & $210.91(3)_{-2.2\%}^{+1.6\%}$ & $202.90(3)_{-1.9\%}^{+1.3\%}$ & $222.41(3)_{-2.5\%}^{+2.2\%}$ & $-3.80$ & $6.20$ & $1.87$ & $1.18$ & $100$\\
\hline
{\fontsize{7.0}{7.0}$W^{+}_{L}W^{-}_{L}$} & $12.99_{-7.4\%}^{+6.1\%}$ & $14.03_{-2.6\%}^{+1.9\%}$ & $13.64_{-2.4\%}^{+1.7\%}$ & $16.46_{-5.7\%}^{+4.7\%}$ & $-2.75$ & $4.08$ & $15.11$ & $0.94$ & $7.4$\\
{\fontsize{7.0}{7.0}$W^{+}_{L}W^{-}_{T}$} & $21.67_{-7.5\%}^{+6.3\%}$ & $24.86_{-2.6\%}^{+1.8\%}$ & $24.28_{-2.5\%}^{+1.7\%}$ & $25.75_{-3.5\%}^{+2.6\%}$ & $-2.32$ & $1.56$ & $3.86$ & $0.50$ & $11.6$\\
{\fontsize{7.0}{7.0}$W^{+}_{T}W^{-}_{L}$} & $22.14_{-7.5\%}^{+6.2\%}$ & $25.56_{-2.6\%}^{+1.8\%}$ & $24.96_{-2.5\%}^{+1.7\%}$ & $26.43_{-3.5\%}^{+2.6\%}$ & $-2.34$ & $1.52$ & $3.75$ & $0.48$ & $11.9$\\
{\fontsize{7.0}{7.0}$W^{+}_{T}W^{-}_{T}$} & $140.44_{-6.0\%}^{+4.8\%}$ & $144.97(2)_{-1.9\%}^{+1.6\%}$ & $138.42(2)_{-1.6\%}^{+1.4\%}$ & $152.95(3)_{-1.9\%}^{+2.3\%}$ & $-4.52$ & $8.32$ & $0.25$ & $1.46$ & $68.8$\\
\hline
{\fontsize{7.0}{7.0}$\text{Interference}$} & $0.90(1)$ & $1.50(4)$ & $1.60(4)$ & $0.81(4)$ & $--$ & $--$ & $--$ & $--$ & $0.4$\\
\hline
\end{tabular}
\caption{\small Unpolarized and doubly polarized cross sections in fb
  calculated in the $VV$ frame for the process $p p \to W^+ W^-\to e^+ \nu_e \mu^- \bar{\nu}_\mu + X$.   
  The statistical uncertainties (in parenthesis) are given on the last
  digits of the central prediction when significant. Seven-point scale
  uncertainty is also provided for the cross sections as sub- and
  superscripts in percent. In the last column the polarization fractions  
  are provided. Taken from \bib{Dao:2023kwc}.}
\label{tab:xs_fr}
\end{bigcenter}
\end{table}

For the second to fourth columns, the cross sections include only the light quark induced processes 
($u$, $d$, $c$, $s$). We will denote the light quark induced processes as $q\bar{q}$, which 
consists of also the quark-gluon and quark-photon induced processes at NLO.
We see that the NLO QCD corrections are moderate because of the jet veto which reduces significantly 
the gluon-quark induced contribution. 
The EW corrections, denoted as $\bar{\delta}_\text{EW}=\Delta\sigma_\text{EW}/\sigma_\text{NLOQCD}$, 
range from $-2\%$ to $-5\%$ for different polarizations. 
In addition, the sub-leading contributions from the loop-induced gluon-gluon fusion ($gg$), photon-photon fusion ($\gamma\gamma$), 
and bottom-antibottom annihilation ($b\bar{b}$) are also included in the $\sigma_\text{all}$, and separately shown 
(relative to the NLOQCD results). 
These small corrections are calculated at LO. 
The most interesting correction is from the $b\bar{b}$ process, where a significant effect of $+15\%$ is found for the 
LL polarization. This large effect is due to the top-quark mass in the $t$-channel propagator. 
We have further investigated this effect at NLO QCD+EW level and presented our results in \cite{Dao:2024ffg}. 
Finally, the polarization fractions ($f_{X} = \sigma_{X}/\sigma_\text{unpol}$) are provided in the last column, 
showing that the LL fraction is about $7\%$, the TT $69\%$, and the LT and TL $12\%$ each. 
The interference is negligible, being $0.4\%$. 

Explaining the hierarchy of the LL, LT, TL, LL cross sections takes a few steps. 
A good way to understand this hierarchy is using the LO results for the on-shell $W^+W^-$ 
production process. The polarized amplitudes for the $q (s_1,q_1) + \bar{q} (s_2,q_2)\to W^+(\lambda_1,p_1) + W^-(\lambda_2,p_2)$ process read
\bea
\mathcal{A}(s_1,s_2,\lambda_1,\lambda_2) = \eps^\mu (\lambda_1,p_1)\eps^\nu (\lambda_2,p_2) \mathcal{M}_{\mu\nu}(s_1,s_2,q_1,q_2),
\eea
where $s_i$, $q_i$ with $i=1,2$ are the helicity indices and momenta of the initial-state quarks, 
$\lambda_i$, $p_i$ the helicity indices and momenta of the $W$ bosons. In the $q\bar{q}$ center of mass system, the momenta read
\begin{align}
q_1^\mu &= (E,0,0,-E),\quad q_2^\mu = (E,0,0,+E);\crn 
p_1^\mu &= (E,-p\sin\theta,0,-p\cos\theta), \quad p_2^\mu = (E,+p\sin\theta,0,+p\cos\theta), 
\end{align}
with $E=(p^2 + M_W^2)^{1/2}$, $p=|{\bf p}_1|=|{\bf p}_2|$.
The polarization vectors of the $W^-$ are
\begin{align}
\eps^\mu (+,p_2) &= \fr{1}{\sqrt{2}} (0,\cos\theta,+i,-\sin\theta), \quad 
\eps^\mu (-,p_2) = \fr{1}{\sqrt{2}} (0,\cos\theta,-i,-\sin\theta); \crn
\eps^\mu (L,p_2) &= \fr{1}{M_W} (p,E\sin\theta,0,E\cos\theta),
\end{align}
which satisfy the orthogonal condition of $\eps(\lambda_2,p_2)\cdot p_2 = 0$ 
and the normalization of $\eps^{*}(\lambda_2,p_2)\cdot\eps(\lambda_2^\prime,p_2) = -\delta_{\lambda_2,\lambda_2^\prime}$.
The polarization vectors of the $W^+$ are obtained by replacing $\theta$ with $(\theta + \pi)$. 

To extract the high energy limit of $E\to \infty$, we write the longitudinal polarization vector in the following form \cite{Willenbrock:1987xz}
\bea
\eps^\mu (L,p_2) = a p_2^\mu + b p_1^\mu.\label{eps_L_decomposed}
\eea
Upon contracting both sides with $p_{2\mu}$ and $p_{1\mu}$; and working in the $WW$ center of mass system where 
$p=\sqrt{s}\beta/2$ with $s=(p_1+p_2)^2 = 4E^2$, $\beta = (1-4M_W^2/s)^{1/2}$; we can easily solve for $a$ and $b$: 
\bea
a = \fr{1+\beta^2}{2\beta M_W}, \quad b = -\fr{2M_W}{\beta s}.
\eea  
In the limit of $E\to \infty$, we have $a \to 1/M_W$ and $b \to 1/E^2$, leading to 
\bea
\eps^\mu (L,p_2) \to \fr{p_2^\mu}{M_W}.
\eea
We can see here an important difference between the logitudinal polarization 
and the transverse ones. While the transverse polarization vectors are independent of the energy, 
the $\eps_L$ is proportional to $E$ in the high energy limit. 

With these ingredients, summing over $s$ and $t$ channel diagrams and neglecting 
the quark masses (the $b\bar{b}$ process is here excluded), one obtains for the partonic cross sections \cite{Willenbrock:1987xz}
\begin{align}
\fr{d\hat{\sigma}_\text{LL}}{dt}&=\fr{\pi\alpha^2}{48s_W^4 \beta^4}\fr{ut-M_W^4}{s^2}\Big{[}\fr{1}{(s-M_Z^2)^2}(b_L^2 + b_R^2)
+ 2\rho^2\fr{1}{s-M_Z^2}\fr{1}{t}b_L
+ \rho^4\fr{1}{t^2}
\Big{]},\\
\fr{d\hat{\sigma}_\text{TL+LT}}{dt}&=\fr{\pi\alpha^2}{48s_W^4\beta^4} \fr{\rho^2}{s^2}
\Big{[}\fr{1}{(s-M_Z^2)^2}[s^2\beta^2-2(ut-M_W^4)](d_L^2+d_R^2) \\
&-4\fr{1}{s-M_Z^2}\fr{1}{t}[s\beta^2(t-M_W^2)+(ut-M_W^4)]d_L \\
&-\fr{4}{t^2}[st\beta^2+\fr{1}{2}(1+\beta^2)(ut-M_W^4)]
\Big{]}, \\
\fr{d\hat{\sigma}_\text{TT}}{dt}&=\fr{\pi\alpha^2}{12s_W^4\beta^4}\fr{ut-M_W^4}{s^4}
\Big{[}
\fr{1}{t^2}(u^2+t^2-2M_W^4) \\
&+ \fr{\rho^4s^2}{8}
  \big{[}\fr{1}{(s-M_Z^2)^2}(d_L^2+d_R^2)
    + 2\fr{1}{s-M_Z^2}\fr{1}{t}d_L
    + \fr{1}{t^2}
  \big{]}
\Big{]},
\end{align}
where $\rho = 2M_W/\sqrt{s}$ originating from the parameter $b$ in \eq{eps_L_decomposed}, $t=(q_1-p_i)^2$ with $p_i$ being 
the momentum of the $W$ with the same sign of the electric charge as the quark, $u=2M_W^2-s-t$ and 
\begin{align}
b_R &= |e_q| \fr{s_W^2}{c_W^2} \beta^2 (3-\beta^2), \quad 
b_L = b_R + \fr{1-2s_W^2}{c_W^2} + 2\rho^2,\\
d_R &= 2 |e_q| \fr{s_W^2}{c_W^2} \beta^2, \quad 
d_L = d_R + \fr{3-4s_W^2}{c_W^2},
\end{align}
with $s_W = \sin\theta_W$, $c_W = \cos\theta_W$, $\theta_W$ being the weak mixing angle, 
$|e_q|=2/3$ for up quarks, $1/3$ for down quarks.
Note that the $L$ and $R$ indices in these coupling parameters refer 
to the helicity of the incoming quarks.
The total partonic cross section is obtained by integrating over $t\in [-s(1+\beta)^2/4,-s(1-\beta)^2/4]$. 
For the protonic cross section, a summation over the light quarks (with 
different weights proportional to the parton distribution functions (PDF)) and 
a further integration over $s\in [4M_W^2, E_\text{CM}^2]$ are performed, 
where $E_\text{CM}=\sqrt{s_{pp}}$ is the proton-proton colliding energy.
Note that the $W^+_T W^-_L$ and $W^+_L W^-_T$ cross sections are equal at the partonic level because of CP invariance, 
but become slightly different at the protonic level due to the PDFs.

We now consider the high energy limit of $E\to \infty$. The partonic cross sections in the region of $|\cos\theta|\ne 1$ read:
\begin{align}
\fr{d\hat{\sigma}_\text{LL}}{dt}&\approx\fr{\pi\alpha^2}{48s_W^4}\fr{-st-t^2}{s^4}(a_L^2 + a_R^2),\\
\fr{d\hat{\sigma}_\text{TL+LT}}{dt}&\approx\fr{\pi\alpha^2}{48s_W^4} \fr{4M_W^2}{s^3}\Big{[}(1+\fr{2t}{s}+\fr{2t^2}{s^2})(c_L^2+c_R^2) 
+\fr{4t}{s}c_L + 4\Big{]},\\
\fr{d\hat{\sigma}_\text{TT}}{dt}&\approx\fr{\pi\alpha^2}{12s_W^4}\fr{-st-t^2}{s^4}(2 + \fr{2s}{t} + \fr{s^2}{t^2}),
\end{align}
with
\begin{align}
a_R &= 2|e_q| \fr{s_W^2}{c_W^2}, \quad 
a_L = a_R + \fr{1-2s_W^2}{c_W^2},\\
c_R &= 2 |e_q| \fr{s_W^2}{c_W^2}, \quad 
c_L = c_R + \fr{3-4s_W^2}{c_W^2}.
\end{align}
A few important remarks are in order. The $1/t$ pole does not occur in the LL and LT cases. 
This means that there is no enhancement in the region of small $|t|$ where $\cos\theta \approx 1$.
This enhancement is however very strong for the TT case. In the limit of $E\to \infty$, 
only the TL and LT cross sections vanish. Comparing the TT and LL cross sections, writing $t\equiv -sx$, we have
\begin{align}
\fr{d\hat{\sigma}_\text{TT}}{d\hat{\sigma}_\text{LL}} \approx \fr{4}{a_L^2 + a_R^2}\Big{[}1+ (1 - \fr{1}{x})^2\Big{]},
\end{align} 
with $x\in [\varepsilon,1]$, $\varepsilon > 0$ ($\varepsilon=0$ when $s=\infty$ and $\cos\theta = 1$). 
The prefactor $4/(a_L^2 + a_R^2)$ is about 
$3.0$ and $4.7$, for up and down quarks respectively, 
explaining why the TT cross section is much larger than the LL one in the high energy region. 
We thus have the hierarchy of $\hat{\sigma}_\text{TT} > \hat{\sigma}_\text{LL} > \hat{\sigma}_\text{LT+TL}$ in the high 
energy limit. 
The nonvanishing value of the LL cross section can be understood from the Goldstone equivalence theorem \cite{Willenbrock:1987xz}. 
This theorem states that, in the high energy limit, the LL cross section is equal to the cross section of the 
$q\bar{q} \to G^+ G^-$ process (where $G^\pm$ are the $W$ Goldstone bosons), 
which does not vanish because of the $s$-channel $\gamma/Z$-exchange diagrams. 

\begin{figure}[th!]
  \centering
  \includegraphics[width=0.40\textwidth]{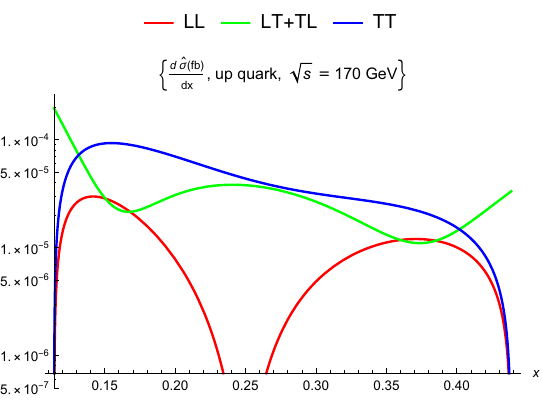}
  \includegraphics[width=0.40\textwidth]{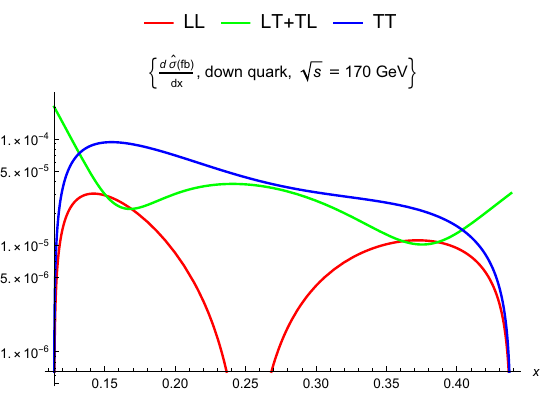}\\
  \includegraphics[width=0.40\textwidth]{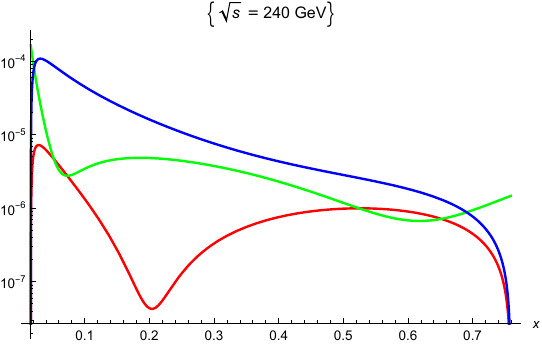}
  \includegraphics[width=0.40\textwidth]{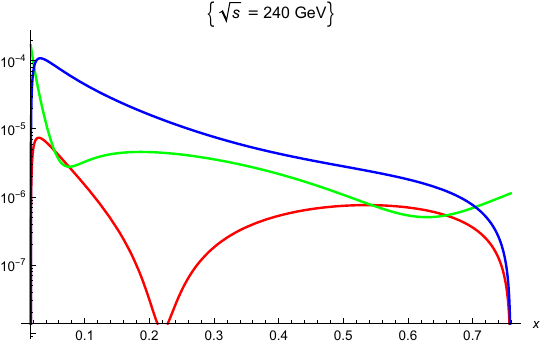}\\
  \includegraphics[width=0.40\textwidth]{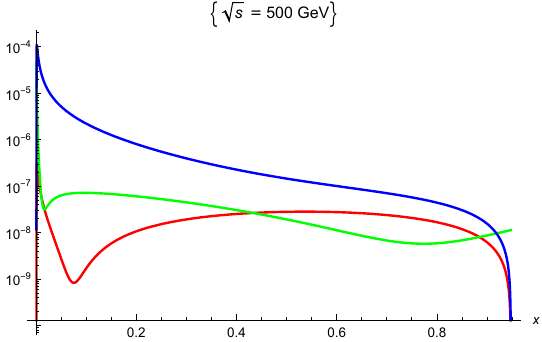}
  \includegraphics[width=0.40\textwidth]{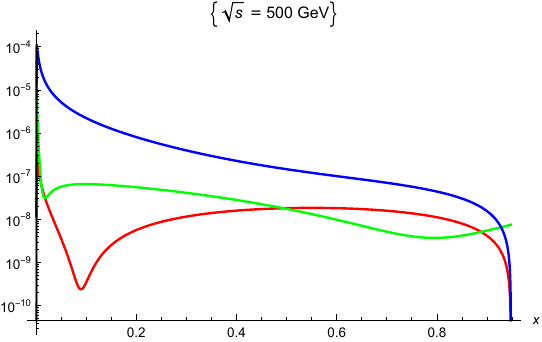}\\
  \includegraphics[width=0.40\textwidth]{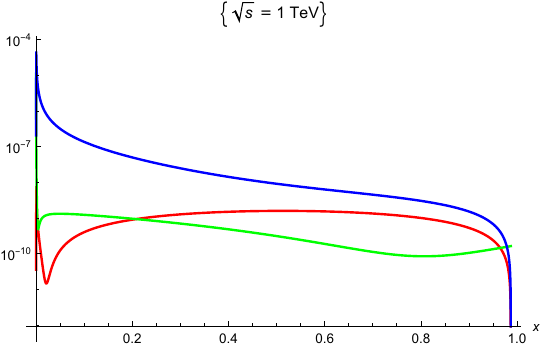}
  \includegraphics[width=0.40\textwidth]{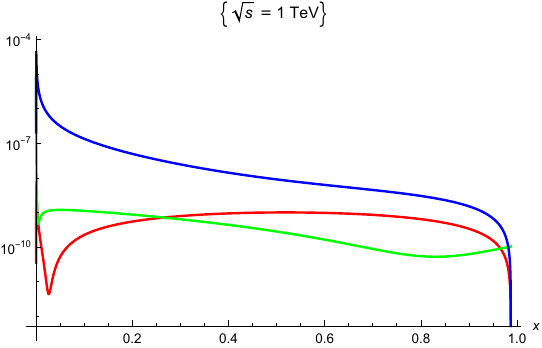}\\
  \includegraphics[width=0.40\textwidth]{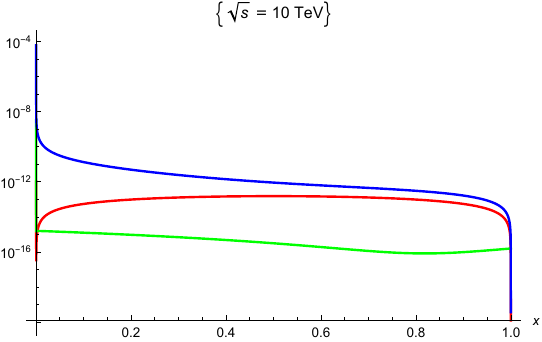}
  \includegraphics[width=0.40\textwidth]{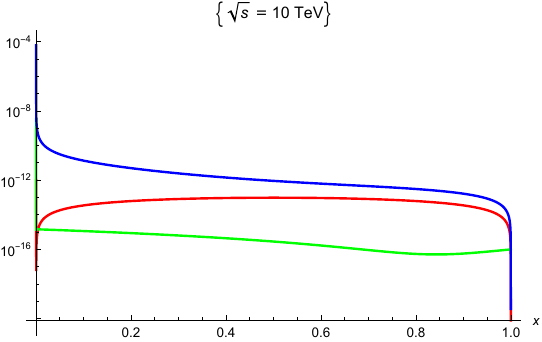}
  \caption{Polarized partonic cross sections as functions of the variable $x=-t/s$ at increasing values 
  of $\sqrt{s}$.}
  \label{fig:plot_ud_x}
\end{figure}

\begin{figure}[th!]
  \centering
  \includegraphics[width=0.48\textwidth]{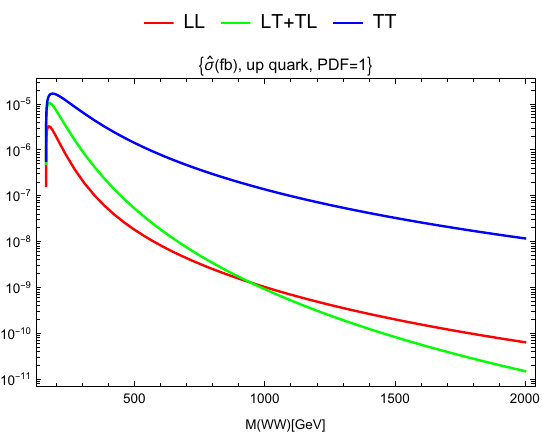}
  \includegraphics[width=0.48\textwidth]{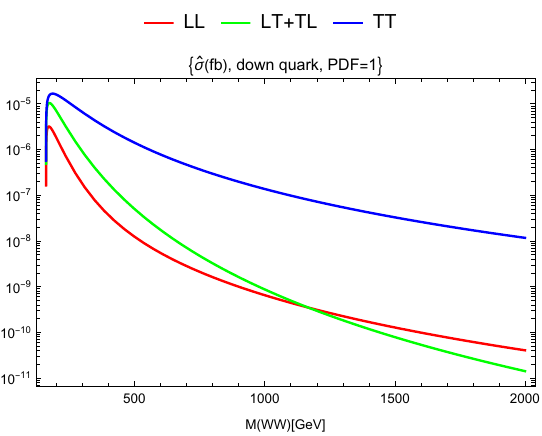}
  \caption{Polarized partonic cross sections as functions of the $WW$ invariant mass.}
  \label{fig:plot_ud_rs}
\end{figure}

\begin{figure}[th!]
  \centering
  \includegraphics[width=0.48\textwidth]{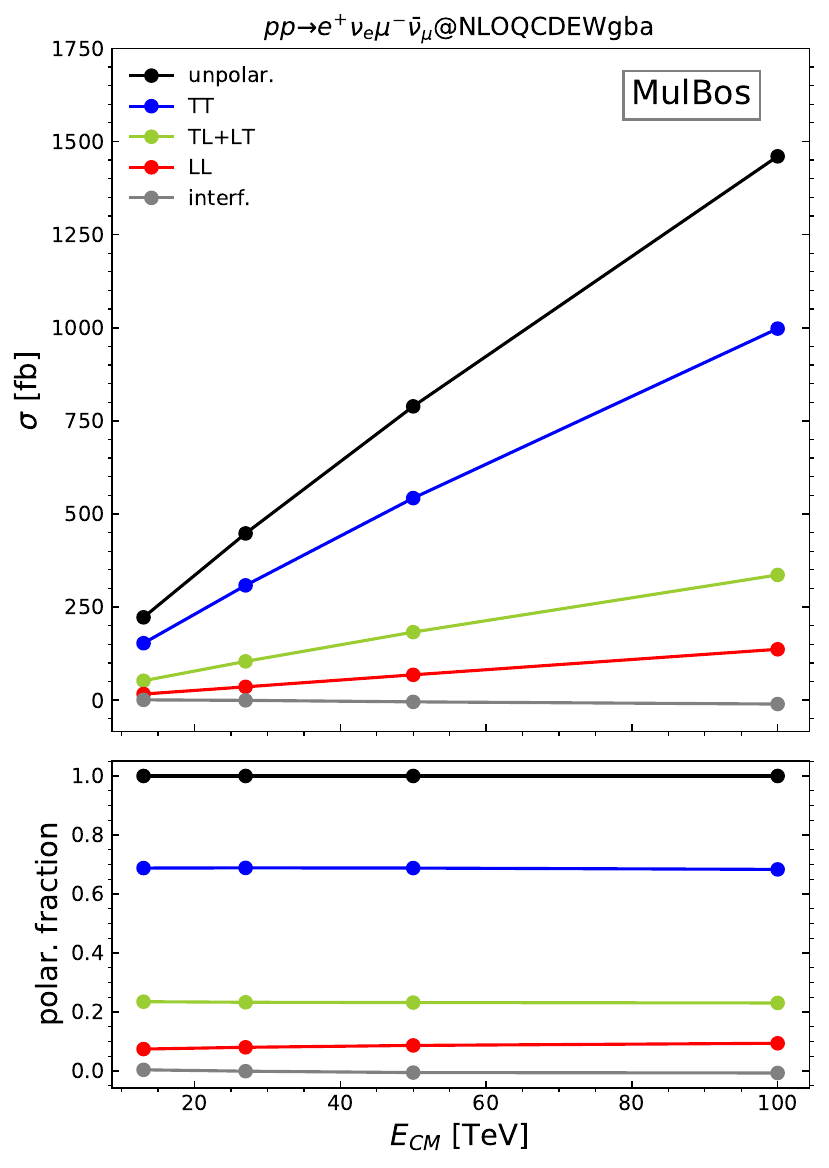}
  \caption{Polarized cross sections as functions of the proton-proton center of mass energy. The 
  corresponding polarization fractions are plotted in the small panel.}
  \label{fig:plot_xs_ecm}
\end{figure}

To understand the observed order of $\sigma_\text{TT} > \sigma_\text{LT} > \sigma_\text{LL}$ as shown in 
\tab{tab:xs_fr} at the LHC we must consider the region of small $s$ where the 
cross sections are largest. In \fig{fig:plot_ud_x}, we show the partonic cross sections as functions 
of $x$ for $\sqrt{s}=$ $170\gev$, $240\gev$, $500\gev$, $1\tev$, and $10\tev$, separately 
for two processes of $u\bar{u}$ and $d\bar{d}$. 
We then integrate over $x$ and show the $M_{WW}=\sqrt{s}$ dependence in 
\fig{fig:plot_ud_rs}. 
Since the protonic cross section is a linear combination of the two plots in 
\fig{fig:plot_ud_rs}, with different weights depending on the value of $\sqrt{s}$, 
it has the same hierarchy of $\hat{\sigma}_\text{TT}$, $\hat{\sigma}_\text{LT}$ and $\hat{\sigma}_\text{LL}$.
Taking into the PDFs, summing over the up and down quark contributions, 
and integrating over $\sqrt{s}$ we then obtain the LO protonic cross sections, which have 
the hierarchy of $\sigma_\text{TT} > \sigma_\text{LT} > \sigma_\text{LL}$ because 
the dominant contribution comes from the region of small $\sqrt{s}$. 
This also explains why the hierarchy does not change as the proton-proton colliding energy increases, 
as can be seen from \fig{fig:plot_xs_ecm} and \tab{tab:xs_ecm}.

\begin{table}[h!]
 \renewcommand{\arraystretch}{1.3}
\begin{bigcenter}
\setlength\tabcolsep{0.03cm}
\fontsize{8.0}{8.0}
\begin{tabular}{|c|c|c|c|c||c|c|c|c|}\hline
 & \multicolumn{4}{c||}{Cross section [fb]} & \multicolumn{4}{c|}{Pol. fraction [\%]} \\
 \hline
$E_\text{CM}\,\text{[TeV]}$  & $13$ & $27$ & $50$ & $100$ & $13$ & $27$ & $50$ & $100$ \\
\hline
{\fontsize{7.0}{7.0}$\text{Unpolarized}$} & $222.41(3)$ & $447.7(2)$ & $788.8(4)$ & $1460(1)$ & $100$ & $100$ & $100$ & $100$ \\
\hline
{\fontsize{7.0}{7.0}$W^{+}_{L}W^{-}_{L}$} & $16.46$ & $35.80(2)$ & $68.08(4)$ & $136.62(8)$ 
& $7.4$ & $8.0$ & $8.6$ & $9.4$ \\
{\fontsize{7.0}{7.0}$W^{+}_{L}W^{-}_{T}$} & $25.75$ & $51.67(3)$ & $90.93(5)$ & $167.7(1)$ 
& $11.6$ & $11.5$ & $11.5$ & $11.5$ \\
{\fontsize{7.0}{7.0}$W^{+}_{T}W^{-}_{L}$} & $26.43$ & $52.50(3)$ & $91.81(5)$ & $168.6(1)$ 
& $11.9$ & $11.7$ & $11.6$ & $11.5$ \\
{\fontsize{7.0}{7.0}$W^{+}_{T}W^{-}_{T}$} & $152.95(3)$ & $308.2(2)$ & $542.5(4)$ & $997.6(8)$ 
& $68.8$ & $68.8$ & $68.8$ & $68.3$ \\
\hline
{\fontsize{7.0}{7.0}$\text{Interference}$} & $0.81(4)$ & $-0.5(3)$ & $-4.5(6)$ & $-10(1)$ 
& $0.4$ & $-0.1$ & $-0.6$ & $-0.7$ \\
\hline
\end{tabular}
\caption{\small Values of the polarized cross sections and the corresponding polarization fractions at $13$, $27$, $50$, $100$ TeV 
of the proton-proton center of mass energy.}
\label{tab:xs_ecm}
\end{bigcenter}
\end{table}
In \fig{fig:plot_xs_ecm} and \tab{tab:xs_ecm} we show the dependence of the polarized cross sections and the corresponding 
polarization fractions on the proton-proton center of mass energy. 
The reference energies of $27$ TeV and $100$ TeV are in accordance with the 
Future Circular Collider Conceptual Design Report \cite{FCC:2018bvk}. 
Since the LT and TL lines almost coincide, they are combined in the plot for better visualization. 
Their individual values are given in \tab{tab:xs_ecm}. 
These results show that, except for the polarization interference, all the polarized cross sections 
scale up linearly with the colliding energy. 
As a consequence, the polarization fractions are flat. 
Most noticeable is the LL fraction, 
which goes very slightly upward, increasing from $7.4\%$ at $13$ TeV to $9.4\%$ at $100$ TeV. 
The polarization interference remains very small, being less than $1\%$, for the whole energy range.

\begin{figure}[th!]
  \centering
  \includegraphics[width=0.48\textwidth]{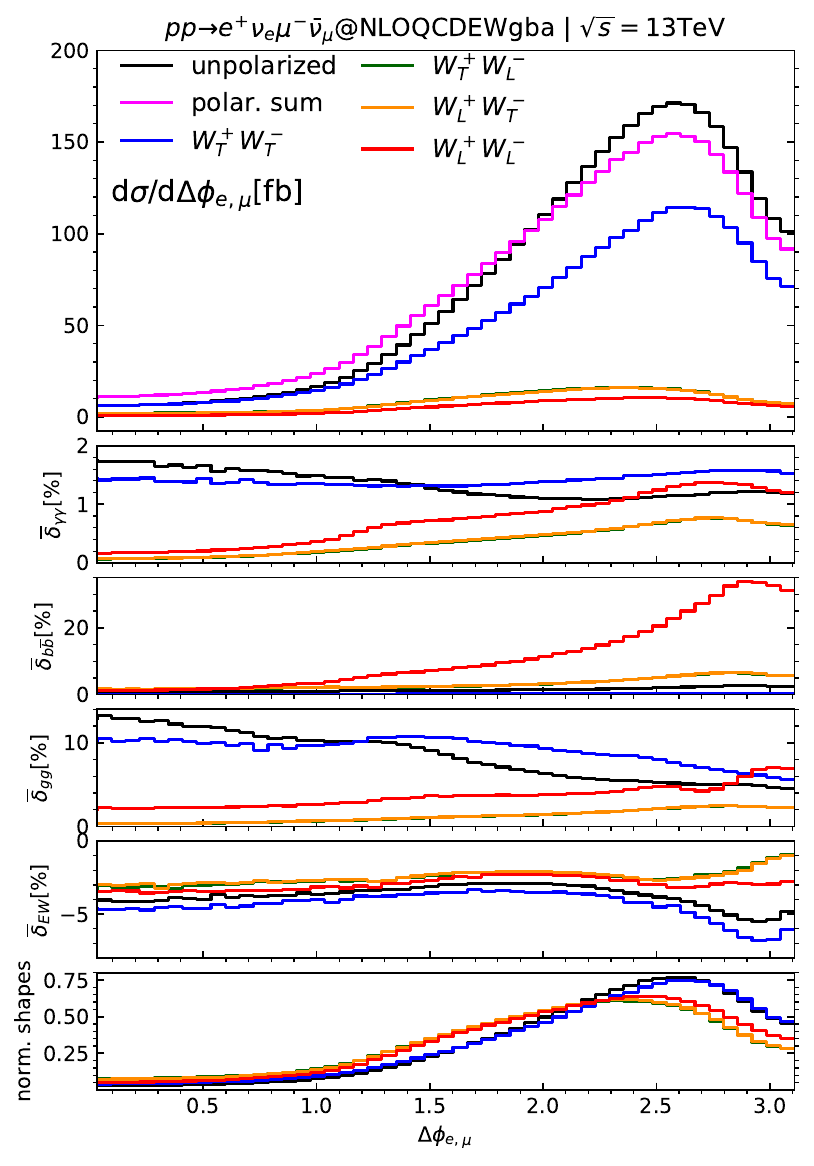}
  \caption{Distributions in the azimuthal-angle separation $\Delta\phi_{e,\mu}$. The big panel
    shows the absolute values of the cross sections including all contributions from 
    the $q\bar{q}$, $gg$, $b\bar{b}$, $\gamma\gamma$ processes. The
    middle panels display the corrections with respect to the NLO QCD $q\bar{q}$ cross sections. 
    The bottom panel shows the
    normalized shapes of the distributions plotted in the top panel. Taken from \bib{Dao:2023kwc}.}
  \label{fig:dist_Delta_phi_e_mu}
\end{figure}

We now discuss kinematic distributions. As an example, we show in \fig{fig:dist_Delta_phi_e_mu} 
the differential distributions in the azimuthal-angle separation between 
the positron and the muon, for individual polarizations.
In this distribution, we see two interesting features. 
First, the interference, which can be seen here as the difference between the unpolarized (black line) 
and the polarization sum (pink line) in the big panel, is quite large, in particular in the 
region of $\Delta\phi \approx 2.5$ where it is about $+12\%$. 
It changes sign at around $\Delta\phi \approx 1.9$, leading to a very small effect in the 
integrated cross section. 
Second, the $\bar{\delta}_{b\bar{b}}$ correction for the LL case increases with the separation, reaching $+20\%$ at 
$\Delta\phi \approx 2.5$. 
Remarkably, this behavior does not change when NLO corrections are included for the bottom-induced processes, as 
can be seen in Fig. 7 of \cite{Dao:2024ffg}.   

Finally, we show in \tab{tab:compare} the comparison between our results (denoted by superscript DL) and \bib{Denner:2023ehn} (denoted DHP) using the input setup of \cite{Denner:2023ehn}, at LO and NLO EW.
\begin{table}[ht!]
 \renewcommand{\arraystretch}{1.3}
\begin{bigcenter}
\setlength\tabcolsep{0.03cm}
\fontsize{7.0}{7.0}
\begin{tabular}{|c|c|c|c||c|c|c|}\hline
  & $\sigma^\text{DL}_\text{LO}\,\text{[fb]}$ & $\sigma^\text{DHP}_\text{LO}\,\text{[fb]}$ & $\Delta_\text{LO}\,\text{[\%]}$ & $\sigma^\text{DL}_\text{NLO}\,\text{[fb]}$ & $\sigma^\text{DHP}_\text{NLO}\,\text{[fb]}$ & $\Delta_\text{NLO}\,\text{[\%]}$\\
\hline
{\fontsize{6.0}{6.0}$\text{unpolar. (DPA)}$} & $245.6(1)$ & $245.79(2)$ & $-0.07$ & $240.56(3)$ & $241.32(2)$ & $-0.3$\\
\hline
{\fontsize{6.0}{6.0}$\text{LL}$} & $18.75(1)$ & $18.752(2)$ & $-0.006$ & $18.497(2)$ & $18.499$ & $-0.01$\\
{\fontsize{6.0}{6.0}$\text{LT}$} & $32.07(2)$ & $32.084(3)$ & $-0.04$ & $31.998(4)$ & $32.032$ & $-0.1$\\
{\fontsize{6.0}{6.0}$\text{TL}$} & $33.21(2)$ & $33.244(5)$ & $-0.09$ & $33.106(4)$ & $33.144$ & $-0.1$\\
{\fontsize{6.0}{6.0}$\text{TT}$} & $182.0(1)$ & $182.17(2)$ & $-0.07$ & $176.93(2)$ & $177.70(2)$ & $-0.4$\\
\hline
\end{tabular}
\caption{\small Comparison between \bib{Dao:2023kwc} (indicated by the superscript DL) and \bib{Denner:2023ehn} (indicated by the superscript DHP) for the $W^+W^-$ process in the $WW$ CM frame at LO and NLO EW levels. 
The difference is defined as $\Delta = (\sigma^\text{DL}-\sigma^\text{DHP})/\sigma^\text{DHP}$. Note: the $\gamma\gamma$ channel is calculated at LO; $b\bar{b}$, $b\gamma$, $gg$ are excluded.
The DHP results are obtained from Table 1 of \bib{Denner:2023ehn} and the $\gamma\gamma$ cross sections (LO and NLO EW) from private communication.
}
\label{tab:compare}
\end{bigcenter}
\end{table}
While the LO agreements are very good, being within 2 standard deviations and with differences smaller 
than 1 per-mille; the NLO EW comparisons are not as good. The largest discrepancy is found for the TT polarization 
at the level of $-0.4\%$. In terms of standard deviation, it is $27\sigma$. 
Compared to the scale uncertainties shown in \tab{tab:xs_fr}, this tiny difference is numerically irrelevant. 
Conceptually, this difference may indicate different implementations of the DPA by the two groups. 
More thorough investigations are currently underway to resolve these discrepancies between the two results.      
\section{Conclusions}
\label{sect:con}
In this contribution, we have presented the NLO QCD+EW results for polarized $W^+W^-$ pairs produced at the LHC with 
a fully leptonic final state. 
A comparison between our results and the ones of \bib{Denner:2023ehn} has been performed, showing good agreements. 
These results complete the polarized calculation of all diboson processes $ZZ$, $WZ$, and $W^+W^-$, being now achieved at the 
NLO QCD+EW level. 
The next step is to implement these calculations into event generators where parton shower and hadronization 
are incorporated, so that experimental colleagues can perform simulations for their measurements.       


\acknowledgments
We would like to thank the organizers of this conference for organizing this wonderful event and the nice atmosphere.  
We are grateful to Ansgar Denner and Giovanni Pelliccioli for helpful discussions and providing us results 
for the comparison with \bib{Denner:2023ehn}.
This research is funded by Phenikaa University under grant number PU2023-1-A-18.


\begin{thebibliography}{10}

\bibitem{OPAL:2000wbs}
{\scshape OPAL} collaboration, G.~Abbiendi et~al., \emph{{Measurement of $W$
  boson polarizations and CP violating triple gauge couplings from $W^{+}
  W^{-}$ production at LEP}},
  \href{https://doi.org/10.1007/s100520100602}{\emph{Eur. Phys. J. C}
  {\bfseries 19} (2001) 229}
  [\href{https://arxiv.org/abs/hep-ex/0009021}{{\ttfamily hep-ex/0009021}}].

\bibitem{DELPHI:2009wdg}
{\scshape DELPHI} collaboration, J.~Abdallah et~al., \emph{{Correlations
  between Polarisation States of W Particles in the Reaction $e^- e^+$ $\to$
  $W^- W^+$ at LEP2 Energies 189 GeV $-$ 209 GeV}},
  \href{https://doi.org/10.1140/epjc/s10052-009-1123-y}{\emph{Eur. Phys. J. C}
  {\bfseries 63} (2009) 611} [\href{https://arxiv.org/abs/0908.1023}{{\ttfamily
  0908.1023}}].

\bibitem{ATLAS:2022oge}
{\scshape ATLAS} collaboration, G.~Aad et~al., \emph{{Observation of gauge
  boson joint-polarisation states in $W^{\pm}Z$ production from pp collisions
  at $\sqrt{s}=13$ TeV with the ATLAS detector}},
  \href{https://doi.org/10.1016/j.physletb.2023.137895}{\emph{Phys. Lett. B}
  {\bfseries 843} (2023) 137895}
  [\href{https://arxiv.org/abs/2211.09435}{{\ttfamily 2211.09435}}].

\bibitem{ATLAS:2024qbd}
{\scshape ATLAS} collaboration, G.~Aad et~al., \emph{{Studies of the energy
  dependence of diboson polarization fractions and the Radiation Amplitude Zero
  effect in WZ production with the ATLAS detector}},
  \href{https://arxiv.org/abs/2402.16365}{{\ttfamily 2402.16365}}.

\bibitem{ATLAS:2023zrv}
{\scshape ATLAS} collaboration, G.~Aad et~al., \emph{{Evidence of pair
  production of longitudinally polarised vector bosons and study of CP
  properties in $ZZ \to 4\ell$ events with the ATLAS detector at $\sqrt{s} =
  13$ TeV}},  \href{https://arxiv.org/abs/2310.04350}{{\ttfamily 2310.04350}}.

\bibitem{CMS:2020etf}
{\scshape CMS} collaboration, A.~M. Sirunyan et~al., \emph{{Measurements of
  production cross sections of polarized same-sign W boson pairs in association
  with two jets in proton-proton collisions at $\sqrt{s} =$ 13 TeV}},
  \href{https://doi.org/10.1016/j.physletb.2020.136018}{\emph{Phys. Lett. B}
  {\bfseries 812} (2021) 136018}
  [\href{https://arxiv.org/abs/2009.09429}{{\ttfamily 2009.09429}}].

\bibitem{Denner:2021csi}
A.~Denner and G.~Pelliccioli, \emph{{NLO EW and QCD corrections to polarized ZZ
  production in the four-charged-lepton channel at the LHC}},
  \href{https://doi.org/10.1007/JHEP10(2021)097}{\emph{JHEP} {\bfseries 10}
  (2021) 097} [\href{https://arxiv.org/abs/2107.06579}{{\ttfamily
  2107.06579}}].

\bibitem{Denner:2020eck}
A.~Denner and G.~Pelliccioli, \emph{{NLO QCD predictions for doubly-polarized
  WZ production at the LHC}},
  \href{https://doi.org/10.1016/j.physletb.2021.136107}{\emph{Phys. Lett. B}
  {\bfseries 814} (2021) 136107}
  [\href{https://arxiv.org/abs/2010.07149}{{\ttfamily 2010.07149}}].

\bibitem{Le:2022lrp}
D.~N. Le and J.~Baglio, \emph{{Doubly-polarized WZ hadronic cross sections at
  NLO QCD + EW accuracy}},
  \href{https://doi.org/10.1140/epjc/s10052-022-10887-9}{\emph{Eur. Phys. J. C}
  {\bfseries 82} (2022) 917}
  [\href{https://arxiv.org/abs/2203.01470}{{\ttfamily 2203.01470}}].

\bibitem{Le:2022ppa}
D.~N. Le, J.~Baglio and T.~N. Dao, \emph{{Doubly-polarized WZ hadronic
  production at NLO QCD+EW: calculation method and further results}},
  \href{https://doi.org/10.1140/epjc/s10052-022-11032-2}{\emph{Eur. Phys. J. C}
  {\bfseries 82} (2022) 1103}
  [\href{https://arxiv.org/abs/2208.09232}{{\ttfamily 2208.09232}}].

\bibitem{Denner:2020bcz}
A.~Denner and G.~Pelliccioli, \emph{{Polarized electroweak bosons in $W^+ W^-$
  production at the LHC including NLO QCD effects}},
  \href{https://doi.org/10.1007/JHEP09(2020)164}{\emph{JHEP} {\bfseries 09}
  (2020) 164} [\href{https://arxiv.org/abs/2006.14867}{{\ttfamily
  2006.14867}}].

\bibitem{Denner:2023ehn}
A.~Denner, C.~Haitz and G.~Pelliccioli, \emph{{NLO EW corrections to
  polarised~W$^+$W$^-$ production and decay at the LHC}},
  \href{https://doi.org/10.1016/j.physletb.2024.138539}{\emph{Phys. Lett. B}
  {\bfseries 850} (2024) 138539}
  [\href{https://arxiv.org/abs/2311.16031}{{\ttfamily 2311.16031}}].

\bibitem{Dao:2023kwc}
T.~N. Dao and D.~N. Le, \emph{{NLO electroweak corrections to doubly-polarized
  $W^+W^-$ production at the LHC}},
  \href{https://doi.org/10.1140/epjc/s10052-024-12579-y}{\emph{Eur. Phys. J. C}
  {\bfseries 84} (2024) 244}
  [\href{https://arxiv.org/abs/2311.17027}{{\ttfamily 2311.17027}}].

\bibitem{Dao:2024ffg}
T.~N. Dao and D.~N. Le, \emph{{Polarized $W^+W^-$ pairs at the LHC: Effects
  from bottom-quark induced processes at NLO QCD+EW}},
  \href{https://arxiv.org/abs/2409.06396}{{\ttfamily 2409.06396}}.

\bibitem{Poncelet:2021jmj}
R.~Poncelet and A.~Popescu, \emph{{NNLO QCD study of polarised $W^{+}W^{-}$
  production at the LHC}},
  \href{https://doi.org/10.1007/JHEP07(2021)023}{\emph{JHEP} {\bfseries 07}
  (2021) 023} [\href{https://arxiv.org/abs/2102.13583}{{\ttfamily
  2102.13583}}].

\bibitem{Denner:2022riz}
A.~Denner, C.~Haitz and G.~Pelliccioli, \emph{{NLO QCD corrections to polarized
  diboson production in semileptonic final states}},
  \href{https://doi.org/10.1103/PhysRevD.107.053004}{\emph{Phys. Rev. D}
  {\bfseries 107} (2023) 053004}
  [\href{https://arxiv.org/abs/2211.09040}{{\ttfamily 2211.09040}}].

\bibitem{Denner:2024ndl}
A.~Denner, D.~Lombardi, S.~L.~P. Chavez and G.~Pelliccioli, \emph{{NLO
  corrections to triple vector-boson production in final states with three
  charged leptons and two jets}},
  \href{https://arxiv.org/abs/2407.21558}{{\ttfamily 2407.21558}}.

\bibitem{Denner:2024xul}
A.~Denner, D.~Lombardi and C.~Schwan, \emph{{Double-pole approximation for
  leading-order semi-leptonic vector-boson scattering at the LHC}},
  \href{https://doi.org/10.1007/JHEP08(2024)146}{\emph{JHEP} {\bfseries 08}
  (2024) 146} [\href{https://arxiv.org/abs/2406.12301}{{\ttfamily
  2406.12301}}].

\bibitem{Hoppe:2023uux}
M.~Hoppe, M.~Sch{\"o}nherr and F.~Siegert, \emph{{Polarised cross sections for
  vector boson production with SHERPA}},
  \href{https://arxiv.org/abs/2310.14803}{{\ttfamily 2310.14803}}.

\bibitem{Pelliccioli:2023zpd}
G.~Pelliccioli and G.~Zanderighi, \emph{{Polarised-boson pairs at the LHC with
  NLOPS accuracy}},  \href{https://arxiv.org/abs/2311.05220}{{\ttfamily
  2311.05220}}.

\bibitem{Javurkova:2024bwa}
M.~Javurkova, R.~Ruiz, R.~C.~L. de~S\'a and J.~Sandesara, \emph{{Polarized ZZ
  pairs in gluon fusion and vector boson fusion at the LHC}},
  \href{https://doi.org/10.1016/j.physletb.2024.138787}{\emph{Phys. Lett. B}
  {\bfseries 855} (2024) 138787}
  [\href{https://arxiv.org/abs/2401.17365}{{\ttfamily 2401.17365}}].

\bibitem{Aeppli:1993cb}
A.~Aeppli, F.~Cuypers and G.~J. van Oldenborgh, \emph{{$\mathcal{O}(\Gamma)$
  corrections to $W$ pair production in $e^+ e^-$ and $\gamma\gamma$
  collisions}}, \href{https://doi.org/10.1016/0370-2693(93)91259-P}{\emph{Phys.
  Lett. B} {\bfseries 314} (1993) 413}
  [\href{https://arxiv.org/abs/hep-ph/9303236}{{\ttfamily hep-ph/9303236}}].

\bibitem{Aeppli:1993rs}
A.~Aeppli, G.~J. van Oldenborgh and D.~Wyler, \emph{{Unstable particles in one
  loop calculations}},
  \href{https://doi.org/10.1016/0550-3213(94)90195-3}{\emph{Nucl. Phys. B}
  {\bfseries 428} (1994) 126}
  [\href{https://arxiv.org/abs/hep-ph/9312212}{{\ttfamily hep-ph/9312212}}].

\bibitem{Denner:2000bj}
A.~Denner, S.~Dittmaier, M.~Roth and D.~Wackeroth, \emph{{Electroweak radiative
  corrections to $e^+ e^- \to W W \to 4\,\text{fermions}$ in double pole
  approximation: The RACOONWW approach}},
  \href{https://doi.org/10.1016/S0550-3213(00)00511-3}{\emph{Nucl.Phys.}
  {\bfseries B587} (2000) 67}
  [\href{https://arxiv.org/abs/hep-ph/0006307}{{\ttfamily hep-ph/0006307}}].

\bibitem{Catani:1996vz}
S.~Catani and M.~Seymour, \emph{{A General algorithm for calculating jet
  cross-sections in NLO QCD}},
  \href{https://doi.org/10.1016/S0550-3213(96)00589-5}{\emph{Nucl.Phys.}
  {\bfseries B485} (1997) 291}
  [\href{https://arxiv.org/abs/hep-ph/9605323}{{\ttfamily hep-ph/9605323}}].

\bibitem{Catani:2002hc}
S.~Catani, S.~Dittmaier, M.~H. Seymour and Z.~Trocsanyi, \emph{{The Dipole
  formalism for next-to-leading order QCD calculations with massive partons}},
  \href{https://doi.org/10.1016/S0550-3213(02)00098-6}{\emph{Nucl. Phys. B}
  {\bfseries 627} (2002) 189}
  [\href{https://arxiv.org/abs/hep-ph/0201036}{{\ttfamily hep-ph/0201036}}].

\bibitem{Dittmaier:1999mb}
S.~Dittmaier, \emph{{A General approach to photon radiation off fermions}},
  \href{https://doi.org/10.1016/S0550-3213(99)00563-5}{\emph{Nucl. Phys.}
  {\bfseries B565} (2000) 69}
  [\href{https://arxiv.org/abs/hep-ph/9904440}{{\ttfamily hep-ph/9904440}}].

\bibitem{Basso:2015gca}
L.~Basso, S.~Dittmaier, A.~Huss and L.~Oggero, \emph{{Techniques for the
  treatment of IR divergences in decay processes at NLO and application to the
  top-quark decay}},
  \href{https://doi.org/10.1140/epjc/s10052-016-3878-2}{\emph{Eur. Phys. J.}
  {\bfseries C76} (2016) 56}
  [\href{https://arxiv.org/abs/1507.04676}{{\ttfamily 1507.04676}}].

\bibitem{Denner:1991kt}
A.~Denner, \emph{{Techniques for calculation of electroweak radiative
  corrections at the one loop level and results for W physics at LEP-200}},
  {\emph{Fortsch.Phys.} {\bfseries 41} (1993) 307}
  [\href{https://arxiv.org/abs/0709.1075}{{\ttfamily 0709.1075}}].

\bibitem{Hahn:2000kx}
T.~Hahn, \emph{{Generating Feynman diagrams and amplitudes with FeynArts 3}},
  \href{https://doi.org/10.1016/S0010-4655(01)00290-9}{\emph{Comput.Phys.Commun.}
  {\bfseries 140} (2001) 418}.

\bibitem{Hahn:1998yk}
T.~Hahn and M.~Perez-Victoria, \emph{{Automatized one-loop calculations in four
  and D dimensions}},
  \href{https://doi.org/10.1016/S0010-4655(98)00173-8}{\emph{Comput. Phys.
  Commun.} {\bfseries 118} (1999) 153}.

\bibitem{Passarino:1978jh}
G.~Passarino and M.~Veltman, \emph{{One Loop Corrections for $e^+$ $e^-$
  Annihilation Into $\mu^+$ $\mu^-$ in the Weinberg Model}},
  \href{https://doi.org/10.1016/0550-3213(79)90234-7}{\emph{Nucl.Phys.}
  {\bfseries B160} (1979) 151}.

\bibitem{'tHooft:1978xw}
G.~'t~Hooft and M.~Veltman, \emph{{Scalar One Loop Integrals}},
  \href{https://doi.org/10.1016/0550-3213(79)90605-9}{\emph{Nucl.Phys.}
  {\bfseries B153} (1979) 365}.

\bibitem{Nhung:2009pm}
D.~T. Nhung and L.~D. Ninh, \emph{{D0C : A code to calculate scalar one-loop
  four-point integrals with complex masses}},
  \href{https://doi.org/10.1016/j.cpc.2009.07.012}{\emph{Comput.Phys.Commun.}
  {\bfseries 180} (2009) 2258}
  [\href{https://arxiv.org/abs/0902.0325}{{\ttfamily 0902.0325}}].

\bibitem{Denner:2010tr}
A.~Denner and S.~Dittmaier, \emph{{Scalar one-loop 4-point integrals}},
  \href{https://doi.org/10.1016/j.nuclphysb.2010.11.002}{\emph{Nucl.Phys.}
  {\bfseries B844} (2011) 199}.

\bibitem{Kawabata:1995th}
S.~Kawabata, \emph{{A New version of the multidimensional integration and event
  generation package BASES/SPRING}},
  \href{https://doi.org/10.1016/0010-4655(95)00028-E}{\emph{Comput. Phys.
  Commun.} {\bfseries 88} (1995) 309}.

\bibitem{Baglio:2024gyp}
J.~Baglio et~al., \emph{{Release note: VBFNLO 3.0}},
  \href{https://doi.org/10.1140/epjc/s10052-024-13336-x}{\emph{Eur. Phys. J. C}
  {\bfseries 84} (2024) 1003}
  [\href{https://arxiv.org/abs/2405.06990}{{\ttfamily 2405.06990}}].

\bibitem{Willenbrock:1987xz}
S.~S.~D. Willenbrock, \emph{{Pair Production of $W$ and $Z$ Bosons and the
  Goldstone Boson Equivalence Theorem}},
  \href{https://doi.org/10.1016/S0003-4916(88)80016-2}{\emph{Annals Phys.}
  {\bfseries 186} (1988) 15}.

\bibitem{FCC:2018bvk}
{\scshape FCC} collaboration, A.~Abada et~al., \emph{{HE-LHC: The High-Energy
  Large Hadron Collider}: {Future Circular Collider Conceptual Design Report
  Volume 4}}, \href{https://doi.org/10.1140/epjst/e2019-900088-6}{\emph{Eur.
  Phys. J. ST} {\bfseries 228} (2019) 1109}.

\end{thebibliography}

\providecommand{\href}[2]{#2}\begingroup\raggedright\endgroup
\end{document}